\documentclass[english]{article}
\usepackage{amsmath}
\usepackage{amssymb}
\usepackage{cite}

\newcommand{\n}[1]{\vec{#1}}

\newcommand{\lapla}{\nabla^{2}}
\newcommand{\derive}[2]{\frac{\partial #1}{\partial #2}}
\newcommand{\derives}[2]{\frac{\partial^{2} #1}{\partial #2^{2}}}
\newcommand{\ederive}[1]{\frac{\partial}{\partial #1}}
\newcommand{\up}[1]{^{\!\!\!\! {^{^{[#1]}}}}}
\newcommand{\varup}[2]{^{\!\!\!\! ^{^{^{[#1]}}} \!\!#2}}
\newcommand{\sjn}{\sum_{j(n)}}
\newcommand{\tot}[2]{\frac{d #1}{d#2}}
\newcommand{\dtot}[2]{\frac{d^{2} #1}{d{#2}^{2}}}
\newcommand{\ttot}[1]{\frac{d #1}{dt}}

\newcommand{\smn}{\sum_{\;m,n_{_{_{\!\!\!\!\!\!\!\!\!\!\!\!\!m\neq
n}}}}}
\newcommand{\sjk}{\sum_{\;j,k_{_{_{_{_{\!\!\!\!\!\!\!\!\!j\neq
k}}}}}}}
\newcommand{\sjkf}{\sum_{\;j,k_{_{_{_{_{\!\!\!\!\!\!\!\!\!j\neq
k}}}}}(free)}}

\newcommand{\bconst}{+}

\makeatletter
\usepackage{color}

\usepackage{graphicx}% Include figure files
\usepackage{dcolumn}% Align table columns on decimal point
\usepackage{bm}% bold math

\usepackage{babel}

\begin{document}
\begin{titlepage}
\thispagestyle{empty}

\bigskip

\begin{center}
\noindent{\Large \textbf
{Deviation of Large Scale Gravitoelectromagnetic Field in Post-Newtonian Approximation}}\\

\vspace{0,5cm}

\noindent{I. C. Jardim${}^{a}$\footnote{e-mail: jardim@fisica.ufc.br}, R. R. Landim ${}^{a}$\footnote{e-mail: renan@fisica.ufc.br}}

\vspace{0,5cm}
 
 {\it ${}^a$Departamento de F\'{\i}sica, Universidade Federal do Cear\'{a}-
Caixa Postal 6030, Campus do Pici, 60455-760, Fortaleza, Cear\'{a}, Brazil. 
 }

\end{center}

\vspace{0.3cm}

\begin{abstract}
In this work a study of the gravity is made using Einstein's equation in the post-Newtonian approach. This is a method to linearise the General Relativity 
indicated to treat non-relativistic objects. It enables us to construct, from metric-independent elements, fields that are governed by equations
similar to the Maxwell's ones in Lorentz gauge. We promediate these equations for matter distributed in local systems, like solar systems or galaxies.
Finally we define the large scale fields for this distribution, which includes terms analogous to electromagnetic case, like polarization, magnetization
and superiors terms.
\end{abstract}
\end{titlepage}

\section{Introduction}
Due to the fact that are known just two long-range interactions, the
gravitation and electromagnetism, came the idea there could be some relationship between them. The discovery of this link could help
for a better understanding of gravity, especially in the sense of this quantization.
Michael Faraday in half of nineteenth century, was, apparently, the first
to imagine that gravitation and electromagnetism have some relationship to each other. Despite his experiments have been a
fundamental importance for the development of electromagnetism, he did not get find any relationship with gravity.
Even the negative results of the experiments in this direction the Faraday's deep conviction about the existence of this connection was not shaken.
When Maxwell published the work in which showed that electric and magnetic phenomena were manifestations of the same
field, the electromagnetic, arose the ambition to unify gravity with these phenomena, until then described by Newtonian theory.
After the reformulation of gravity by Einstein the attempts to unify the long-range interactions have taken a new direction.
In 1918, Weyl \cite{Weyl:1918ib} published a paper which proposed to modify the connection of general relativity by another in which the electromagnetic field
would be contained, in the mathematical point of view meant leaving the Riemannian geometry.
Another attempt to modify the geometry in order to unify these two interactions was proposed by Klein and Kaluza, and consisted of
maintain the Riemannian geometry, but include an extra spatial dimension \cite{Kaluza:1921tu, Klein:1926tv}. The degrees of freedom of gravity in more dimensions, after a dimensional reduction,
would fall in the usual gravity and electromagnetism.
Einstein himself was convinced that this relationship should exist and devoted the last years of his life to the search of the unified field theory,
that is still the subject of countless works \cite{Smolin:2007rx, Dijkgraaf:2004te, Weinberg:2000sk, Lisi:2007gv, Weinberg:1992nd, Laughlin:2000wr}.
Despite all failed attempts of unification made by great names of science persists the belief that this connection exists.

Historically the attempt to establish an analogy between electromagnetism and gravitation began in 1893, when Heaviside  \cite{o.heaviside-electrician31}
investigated how the energy propagates in a gravitational field. He proposed a gravitational Poynting vector which contained magnetic components
of gravitation. In Newton's theory of gravitation a problem is the absence of freedom of the gravitational field, which is fully defined by a single
scalar field, while the electromagnetism by a scalar and a vector field.
We can associate the Newtonian gravitational potential to electromagnetic scalar potential, but to proceed with an analogy
between both theories we need a vector potential for gravity. In Einstein's theory, the gravitation gains more freedom because now is determined by the 10 independent components of the 
metric tensor. In General Relativity is the gravity who has a lot of freedom in relation to electromagnetism. Another problem raised
by the theory of General Relativity is that its equations are nonlinear while
Maxwell's are. So, to create an analogy between them it is necessary to linearise General Relativity and reduce the degrees of freedom of the metric.
Forward was the first to write the Einstein's equations  in a linearised Maxwell-type structure, and proposed experiments to detect the gravitomagnetic fields  \cite{rl.forward-pire49, Forward:1963qb}.
Braginski \cite{PhysRevD.15.2047} reached similar equations using the parametrized post-Newtonian approximation \cite{w.misner-grav}, 
while Campbell left in a tensorial form  \cite{Campbell:1970ww, wb.campbell-ajp44}.
Experiments in superconductors was recently proposed by Li \cite{n.li-prd43, n.li-prb46, n.li-pc281} following the corrections introduced by Ross \cite{dk.ross-jpa16} and deWitt \cite{bs.dewitt-prl16} in London equation to include the gravitomagnetic fields.

In this paper we use the post-Newtonian approximation one order beyond the Newtonian limit in order to define the gravitoelectromagnetic field with independent terms of the metric.
The found equations are used to calculate the effective fields in large scale, considering the matter distributed in local systems. This systems can be planets orbiting a star or galaxies, depending on the desired scale.
The post-Newtonian approximation is satisfactory in these cases since the speed of revolution are much smaller than the light \cite{Weldrake:2002ri, 1978ApJ...225L.107R, 1977ApJ...217L...1R, 1978AJ.....83.1564T, 1979ApJ...230...35R, 2001ARA&A..39..137S}.
The deviation of effective fields is made in an analogous way which was made by Russakoff in electromagnetic case \cite{g.russakoff-ajp38}.
In this study we expand the contribution of the bodies bound to local systems in power series in relation to the size of these systems. Finally, we collect the terms analogues to polarization, magnetization and superiors terms, such as quadrupole, to define the effective fields in large scale.

This paper is organized as follows: In Sec. II we make a review about the post-Newtonian approximation, in particular the expansion of the metric and the approximate field equations.
We also expand the energy-momentum tensor used in this work and obtain general solutions.
In Sec. III we write the equations in the form of Maxwell's equations of electromagnetism, as well as we define the gravitoelectromagnetic field with degrees of freedom of the metric.
In Sec. IV we apply the Maxwell-Einstein's equations for local systems and calculate the promediation of the sources making a multipole expansion in terms of typical lengths of these systems.
In Sec. V we define the effective fields in large scale in an analogous way to electromagnetism and write the equations obeyed by these fields. In the last section we discuss the conclusion of this work and present possibles perspectives.

\section{ The post-Newtonian approximation}
To linearise the Einstein's equation we use the post-Newtonian approximation, which is a method suitable for objects moving at
low speeds compared to the speed of light. This approach is recommended to the problem we want to study, which is the
bodies bounded in  solar-type systems. This is because post-Newtonian approximation inserts corrections in Newtonian mechanics, and this already sufficiently precise that even situations for sending space probes.
This approach consists in expand the components of the metric tensor and of the energy-momentum tensor in powers of the average speed \cite{s.weinberg-gc}, i.e., we expand the metric as
\begin{eqnarray}
g_{00} &=& -1 +g_{00}\up{2} +g_{00}\up{4} +\cdots ,
\\ g_{ij} &=& \delta_{ij} +g_{ij}\up{2} +g_{ij}\up{4} +\cdots ,
\\ g_{i0} &=& g_{i0}\up{3} +g_{i0}\up{5} +\cdots,
\end{eqnarray}
where the symbol $g_{\mu\nu}\up{N}$ denotes the terms of $g_{\mu\nu}$ in order $N$ of typical speed $\bar{v}$.
In this paper we consider one term higher than the Newtonian theory, i.e., we will consider up to fourth order in $g_{00}$,
third order in $ g_{i0} $ and second order in $ g_{ij} $. Using the above expansion of the metric in Einstein's field equation we get the set of equations
\begin{eqnarray}
\nabla^{2}g_{00}\up{2} &=& -8\pi GT\varup{0}{00} ,\label{aprox1.1}
\\ \lapla g_{00}\up{4} &=& \frac{\partial^{2}g_{00}\up{2}}{\partial t^{2}} +
g_{ij}\up{2}\frac{\partial^{2}g_{00}\up{2}}{\partial x^{i}\partial
x^{j}} -
\derive{g_{00}\up{2}}{x^{i}}.\derive{g_{00}\up{2}}{x^{i}} - \nonumber
\\&&-8\pi G[T\varup{2}{00} -2g_{00}\up{2}T\varup{0}{00} +T\varup{2}{ii}], \label{aprox1.2}
\\ \lapla g_{i0}\up{3} &=& 16\pi GT\varup{1}{i0} ,\label{aprox1.3}
\\ \lapla g_{ij}\up{2} &=& -8\pi G\delta_{ij}T\varup{0}{00}, \label{aprox1.4}
\end{eqnarray}
where $ T\varup{N}{\mu \nu} $ denotes the term of $ T^{\mu \nu} $ in order of $\bar{m}\bar{v}^{N}/\bar{r}^{3} $, where $\bar{m}$ and $\bar{r}$ are the typical values of mass and distance.
The above equations been simplified by a convenient choice of coordinate system, which obeys the Einstein's gauge, defined by 
\begin{equation}\label{GE}
 \Gamma^{\lambda} \equiv g^{\mu\nu}\Gamma^{\lambda}_{\mu\nu} = 0.
\end{equation}
Since we are interested in this work to apply this formalism to the universe in large scale, we can approximate the celestial bodies
by point objects. In this context we use the energy-momentum tensor of particles that interact gravitationally and, sometimes, by localized collisions \cite{s.weinberg-gc}
\begin{small}
\begin{equation}\label{Tfull}
 T^{\mu\nu}(\vec{x},t) = g^{-1/2}\sum_{n}m_{n}\frac{dx^{\mu}_{n}}{dt}\frac{dx^{\nu}_{n}}{dt}\left(\frac{d\tau_{n}}{dt}\right)^{-1}\delta(\n{x} - \n{x}_{n}). 
\end{equation}
\end{small}
Using the expansion of the metric we obtain the terms needed for the present work
\begin{eqnarray}
T\varup{0}{00} &=& \rho_{0} ,\label{tmini1}
\\ T\varup{2}{00} &=& \rho_{k} +\rho_{0}\left(g_{00}\up{2} -\frac{1}{2}g\up{2}_{ii}\right) ,\label{tmini2}
\\ T\varup{1}{0i} &=& j^{i}_{0} ,\label{tmini3}
\\ T\varup{2}{ij} &=& \sum_{n}m_{n}v^{i}_{n}v^{j}_{n}\delta(\vec{x} -\vec{x}_{n}(t)) ,\label{tmini4}
\end{eqnarray}
where
\begin{eqnarray}
&&\rho_{0} = \sum_{n}m_{n}\delta(\vec{x} -\vec{x}_{n}(t)),
\\&& j^{i}_{0} = \rho_{0}v^{i} = \sum_{n}m_{n}v^{i}_{n}\delta(\vec{x} -\vec{x}_{n}(t)),
\\&&\rho_{k} = \frac{1}{2}\sum_{n}m_{n}v_{n}^{2}\delta(\vec{x} -\vec{x}_{n}(t)),
\end{eqnarray}
are the density of matter, the current of matter and the density of kinetic energy, respectively. 
With the components of the energy-momentum tensor we can solve the approximated field equations. Considering that the fields vanish at infinity and  using the Green's theorem \cite{jackson}  we can write the solutions in the form
\begin{eqnarray}
&& g_{00}\up{2} = -2\phi_{n},
\\&& g_{00}\up{4} = -2\phi_{n}^{2} -2\psi,
\\&& g_{ij}\up{2} =  -2\delta_{ij}\phi_{n},
\\&& g_{i0}\up{3} = -4\chi_{i}
\end{eqnarray}
where
\begin{eqnarray}
\phi_{n} &=& G\int d^{3}x'\frac{\rho_{0}(\vec{x}',t)}{|\vec{x} - \vec{x}'|}
\\\chi_{i} &=&  G\int d^{3}x'\frac{j_{0i}(\vec{x}',t)}{|\vec{x} - \vec{x}'|}
\\ \psi &=& -\frac{1}{4\pi}\int\frac{d^{3}x'}{|\vec{x} - \vec{x}'|}\derives{\phi_{n}(\vec{x}',t)}{t} - \nonumber
\\&&-G\int\frac{d^{3}x'}{|\vec{x} - \vec{x}'|}\left[\rho_{g}(\vec{x}',t) +3\rho_{k}(\vec{x}',t)\right],
\end{eqnarray}
with $\rho_{g}$ been the gravitational energy density
\begin{equation}
\rho_{g} = \sum_{n}m_{n}\phi_{n}\delta^{3}(\n{x} - \n{x_{n}}).
\end{equation}
The solution to $g_{00}\up{2}$ was already expected since $ \phi_{n} $ is the Newtonian gravitational potential, exactly the non-relativistic limit of General Relativity.

\section{The Maxwell-Einstein's equations and the gravitoelectromagnetic field}
In the previous section we find general solutions to the approximate equations. In this section we write these equations in a form analogous to Maxwell's equations of electromagnetism.
Neglecting terms of orders higher than those considered in this work, we can write the metric in the form
\begin{eqnarray} 
g_{00} &=&  -1 -2\left(\phi +\phi^{2}\right),
\\ g_{ij} &=& \left[ 1 -2\phi\right] \delta_{ij},
\\ g_{i0} &=& 4A_{i},
\end{eqnarray}
where we defined the fields
\begin{eqnarray}
&&\phi \equiv \phi_{n} + \psi,
\\&& A_{i} = - \chi_{i}.
\end{eqnarray}
In this approach the metric is uniquely determined by four degrees of freedom, a scalar field $ \phi $ and a vector field $ \vec{A} $, which satisfy the following equations
\begin{eqnarray}
\lapla \phi &=& \derives{\phi_{n}}{t}  +4\pi G\left[\rho_{0} + \rho_{g} +3\rho_{k}\right], \label{laplaphi}
\\ \lapla \vec{A} &=&  4\pi G \vec{j}_{0}.
\end{eqnarray} 
In this approach the gravity has the same number of degrees of freedom that the electromagnetic field, so that we can identify them as
four components of the electromagnetic quadrivector $ A^{\mu} $. Another indication that we are in a correct way to build an analogy between
gravitation and electromagnetism is that the not trivially satisfied component of harmonic coordinates fixation (\ref{GE}), gives, in this approximation, the condition
\begin{equation}\label{gaugelorentz}
\derive{\phi}{t} + \nabla.\vec{A} = 0.
\end{equation}
That, analogous to electromagnetism, is the Lorentz gauge. This condition allows us to define the gravitoelectric and gravitomagnetic fields analogously to electromagnetism, i.e.
\begin{eqnarray}
\vec{E}_{g} &\equiv& -\nabla\phi -\derive{\vec{A}}{t} ,\label{gravitoeletrico}
\\ \vec{B}_{g} &\equiv& \nabla\times\vec{A} .\label{gravitomagnetico}
\end{eqnarray}
With the same calculation used in electromagnetism we obtain, in this approximation, the Maxwell-Einstein's equations
\begin{eqnarray}
\nabla.\vec{E}_{g} &=& -\frac{1}{\varepsilon_{g}}\left[\rho_{0} + \rho_{g} +3\rho_{k}\right] ,\label{m-e1}
\\ \nabla.\vec{B}_{g} &=& 0 ,\label{m-e2}
\\ \nabla\times\vec{E}_{g} &=& -\derive{\vec{B}_{g}}{t} ,\label{m-e3}
\\ \nabla\times\vec{B}_{g} &=& -\mu_{g}\vec{j}_{0} + \derive{\vec{E}_{g}}{t}, \label{m-e4}
\end{eqnarray} 
where we define the constants
\begin{eqnarray}
 \varepsilon_{g}^{-1} = \mu_{g} \equiv 4\pi G.
\end{eqnarray}
Analogously we can obtain the equations for the fields separately, which are
\begin{eqnarray}
&& \lapla\vec{E}_{g} -\derives{\vec{E}_{g}}{t} = \frac{1}{\varepsilon_{g}}\left[\derive{\vec{j}_{0}}{t} +\nabla\left(\rho_{0} +\rho_{g} +3\rho_{k}\right)\right] .\label{ondaE}
\\&&\lapla\vec{B}_{g} = \mu_{g}\nabla\times\vec{j}_{0} .\label{ondaB}
 \end{eqnarray}
In the last equation was neglected the term with second derivatives of $ \vec{B}_{g} $ as that term is higher order them the approximation used here.
We can use the Green's theorem to obtain the solution of (\ref{ondaE}) and (\ref{ondaB}) which are
\begin{eqnarray}
\vec{E}_{g} &=& -G\int\frac{d^{3}x'}{|\vec{x} -\vec{x}'|}\left[\derive{\vec{j}_{0}(\vec{x}',t')}{t'} +\nabla'\left[\rho_{0}(\vec{x}',t')  +\right.\right. \nonumber
\\&&+\rho_{g}(\vec{x}',t') +\left.\left.3\rho_{k}(\vec{x}',t')\right]\right]_{RET} , 
\\ \vec{B}_{g} &=& -G\int d^{3}x'\frac{\nabla'\times\vec{j}_{0}(\vec{x}',t')}{|\vec{x} -\vec{x}'|} ,
\end{eqnarray}
where the bracket $ [\; \;]_{RET}$ means that $ t '$ is the retarded time, $t' = t - |\vec{x} -\vec{x}'|$. 

The differences between these equations to Maxwell's is the negative signals in sources of gravitoelectric, and gravitomagnetic, fields which
gives due to the fact that gravity is attractive while electromagnetism is repulsive for equal charges. Another difference is that the source of
gravitoelectric field is not only the matter density but contains terms proportional to the density of gravitational energy and kinetic energy, which comes from the fact that not only the mass interacts gravitationally but also the energy.
Although we had show be possible to construct fields analogous to electromagnetic with the independent terms of the metric the analogy can not be taken later on.
This is because is not possible write the equation of motion in the form of the Lorentz force. The geodesic equation, in this approximation may be written as
\begin{eqnarray}
\dtot{\vec{x}}{t} &=& -\left[\
\nabla(\phi +2\phi^{2} ) +4\derive{\vec{A}}{t} \right] + 4\vec{v}\times\left(\nabla\times\vec{A}\right) +\nonumber
\\&&+3\vec{v}\derive{\phi}{t} +4\vec{v}(\vec{v}.\nabla)\phi -v^{2}\nabla\phi,
\end{eqnarray}
which has terms in direction of velocity, and bilinear terms, which does not allow us to write it in the form of Lorentz force. 

\section{Deviation of large scale equations}
\begin{figure}
 \centering
 \includegraphics[scale=0.6]{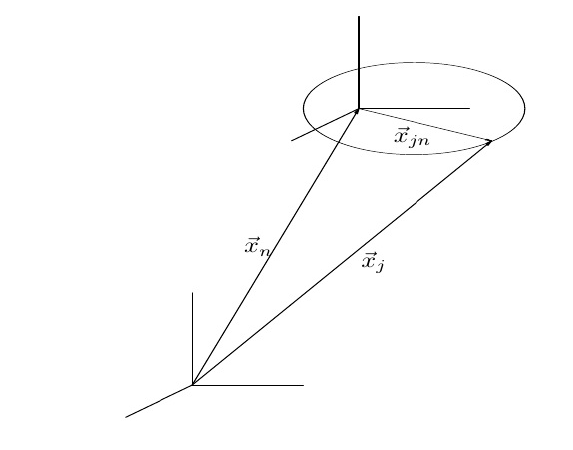}
 % coord.pdf: 842x595 pixel, 72dpi, 29.70x20.99 cm, bb=0 0 842 595
 \caption{Local coordinate systems. It is used to write the position of the bodies connected to local systems, where $ \vec{x}_{n} $ orients the center of mass of $n$-th local system and $\vec{x}_{j}$ the $j$-th bounded body, respecting the relation 
 $ \vec{x}_{j} = \vec {x}_{n} + \vec{x}_{jn}$.} 
 \label{fig: jjjj}
\end{figure}

In this section we derive the effective Maxwell-Einstein's equations in large-scale, analogously to the electromagnetic case \cite{g.russakoff-ajp38,jackson}.
In electromagnetic case is made a power series expansion on the molecules internal distances, considered much smaller than the scale of interest.
In this work we will do a similar procedure, but we will expand on the internal distances of local systems. These systems can be solar systems or galaxies,
since they are similar bounded structures, depending on the distances of interest. For this we will use a coordinate system similar to that used by Russakoff \cite{g.russakoff-ajp38}, 
which localize the bodies bounded in local systems in relation to the mass center of the system, as illustrated in FIG. \ref{fig: jjjj}.
Considering that the bodies involved produces fields that obey the equations
\begin{eqnarray}
\nabla.\vec{e}_{g} &=& -\frac{1}{\varepsilon_{g}}\left[\rho_{0}  +3\rho_{k} \bconst\rho_{g}\right] ,
\\ \nabla.\vec{b}_{g} &=& 0 ,
\\ \nabla\times\vec{e}_{g} &=& -\derive{\vec{b}_{g}}{t} ,
\\ \nabla\times\vec{b}_{g} &=& -\mu_{g}\vec{j}_{0} + \derive{\vec{e}_{g}}{t} ,
\end{eqnarray} 
we must compute the spatial average of the amounts above to obtain the effective equations for larger scales.
The spatial average of a function $ F (\vec{x}, t) $ in relation to a function test $ w (\vec{x}) $ is defined as
\begin{equation}\label{promed}
\left< F(\vec{x},t) \right> \equiv \int d^{3}x'w(\vec{x}')F(\vec{x} -\vec{x}',t) , 
\end{equation} 
where $ w(\vec{x}) $ is a real function, non-zero in a neighborhood of $ \vec{x} = 0 $ and normalized to unity in the whole space\cite{g.russakoff-ajp38}.
The test function need not be specified in detail, all that is needed are general properties of continuity and smooth variation. This allows the rapid convergence of the Taylor series development over distances of the local systems size, i.e., that can write
\begin{eqnarray}\label{exp}
 w(\vec{x} -\vec{x}_{n} -\vec{x}_{jn}) &\approx& w(\vec{x} -\vec{x}_{n})  -\vec{x}_{jn}.\nabla w(\vec{x} -\vec{x}_{n}) + \nonumber
 \\&&+\frac{1}{2}\nabla.\nabla.\left[\vec{x}_{jn}\vec{x}_{jn}w(\vec{x} -\vec{x}_{n})\right].
\end{eqnarray}

Respected the condition for the test function we can define the large-scale gravitoelectric and gravitomagnetic fields, $ \vec{E}_{g} $ and $ \vec{B}_{g} $, as the averages of fields $ \vec{e}_{g} $ and $ \vec{b}_{g} $, i.e.,
\begin{equation}
 \vec{E}_{g} \equiv <\vec{e}_{g}>\;\;\; \mbox{e} \;\;\;\vec{B}_{g} \equiv <\vec{b}_{g}>,
\end{equation}
so, the promediate Maxwell-Einstein's equations are 
\begin{eqnarray}
\nabla.\vec{E}_{g} &=& -\frac{1}{\varepsilon_{g}}\left[\left<\rho_{0}\right>  +3\left<\rho_{k}\right> \bconst\left<\rho_{g}\right>\right] ,
\\ \nabla.\vec{B}_{g} &=& 0 ,
\\ \nabla\times\vec{E}_{g} &=& -\derive{\vec{B}_{g}}{t} ,
\\ \nabla\times\vec{B}_{g} &=& -\mu_{g}\left<\vec{j}_{0}\right> + \derive{\vec{E}_{g}}{t} .
\end{eqnarray} 
To find the desired equations we need to calculate the spatial averages of $ \rho_{0} $, $ \rho_{g} $, $ \rho_{k} $ and $ \vec {j}_{0} $, that we will calculate below.

\section*{Calculation of spatial average of $\rho_{0}$}
Considering that in large-scale the involved bodies can be taken as point, in the coordinate system shown in FIG. \ref{fig: jjjj}, the density of matter, $ \rho_{0}$, can be written as
\begin{equation}
\rho_{0} = \sum_{j(free)}m_{j}\delta(\vec{x} -\vec{x}_{j})
+\sum_{n}\sum_{j(n)}m_{j}\delta(\vec{x} -\vec{x}_{n}
-\vec{x}_{jn}) ,\label{medro}
\end{equation}
where $\vec{x}_{n}$ orients the center of mass of the $ n $-th local system. The first term of $ \rho_{0} $ is the contribution of free bodies
while the second term represents the contribution of these systems. The average of density of matter of $n$-th local system is, using the definition (\ref{promed}), given by 
\begin{equation}
<\rho_{0}>_{n} = \sum_{j(n)}m_{j}w(\vec{x} -\vec{x}_{n} -\vec{x}_{jn}) .
\end{equation}
As we are considering the dimensions of these local systems much smaller than the dimensions considered, we can use the expansion (\ref{exp}) to write
\begin{eqnarray}
<\rho_{0}>_{n} &\approx& \sjn m_{j}w(\vec{x} -\vec{x}_{n}) -\sjn(m_{j}\vec{x}_{jn}).\nabla w(\vec{x} -\vec{x}_{n})+ \nonumber
\\ &&+\frac{1}{2}\sjn m_{j}(\vec{x}_{jn}.\nabla)[\vec{x}_{jn}.\nabla w(\vec{x} -\vec{x}_{n})].
\end{eqnarray}
As $ \vec{x}_{n} $ guides the center of mass of the $n$-th local system, then $\sjn m_{j}\vec{x}_{jn} = \vec{0}$. So we can write
\begin{equation}
<\rho_{0}>_{n} \approx <m_{n}\delta(\vec{x} -\vec{x}_{n})> +\nabla.\left[\nabla.<\textbf{q}_{n}\delta(\vec{x} -\vec{x}_{n})>\right],
\end{equation}
where
\begin{equation}
\textbf{q}_{n} \equiv \frac{1}{2}\sjn m_{j}\vec{x}_{jn}\vec{x}_{jn} ,
\end{equation}
is the quadrupole moment of the $n$-th local system. Finally, according to (\ref{medro}), we can write the average of density of matter of the form
\begin{equation}
<\rho_{0}> \approx \rho_{f} +\nabla.\left[\nabla.\textbf{Q}\right] ,\label{rofinal}
\end{equation}
where we define the density of free matter and the quadrupole momentum
\begin{equation}
\rho_{f} \equiv \left<\sum_{j(free)}m_{j}\delta(\vec{x} -\vec{x}_{j})\right> +\left<\sum_{n}m_{n}\delta(\vec{x} -\vec{x}_{n})\right> ,
\end{equation}
\begin{equation}
\textbf{Q} \equiv \left<\sum_{n}\textbf{q}_{n}\delta(\vec{x} -\vec{x}_{n})\right> .
\end{equation}
The result (\ref{rofinal}) is similar to obtained in electromagnetic case with electric charge exchanged by mass. This result has not analogous term to dipole momentum as that term represents the sum of the momentum of each local system, which vanish due to reference system taken on center of mass.

\section*{Calculation of spatial average of $\vec{j}_{0}$}
Analogously to what was done in calculation of average of $ \rho_{0} $ we can write, using the same coordinate system, $ \vec{j}_{0} $ as follows
\begin{eqnarray}
\vec{j}_{0} &=& \sum_{j(free)}m_{j}\vec{v}_{j}\delta(\vec{x} -\vec{x}_{j})+ \nonumber
\\&&+\sum_{n}\sum_{j(n)}m_{j}(\vec{v}_{n} +\vec{v}_{jn})\delta(\vec{x} -\vec{x}_{n} -\vec{x}_{jn}) ,\label{medj0}
\end{eqnarray}
where $\vec{v}_{j} = \ttot{\vec{x}_{j}}$. Again, the first term is the contribution of free mass while the second is the contribution of the bounded bodies. We will calculate separately the average of bounded term which we can expand in series.
The average of current matter of the $n$-th local system is given by
\begin{eqnarray}
<\vec{j}_{0}>_{n} &=& \left<\sum_{j(n)}m_{j}(\vec{v}_{n} +\vec{v}_{jn})\delta(\vec{x} -\vec{x}_{n} -\vec{x}_{jn})\right> \nonumber
\\ &\approx&  m_{n}\vec{v}_{n}w(\vec{x} -\vec{x}_{n}) +\vec{v}_{n}\nabla.\nabla.[\textbf{q}_{n}w(\vec{x} -\vec{x}_{n})]- \nonumber
\\&&-\sjn(m_{j}\vec{v}_{jn}\vec{x}_{jn}).\nabla w(\vec{x} -\vec{x}_{n}) ,\label{j0}
\end{eqnarray}
where was neglected terms proportional to $\vec{v}_{jn}\vec{x}_{jn}\vec{x}_{jn}$.
To simplify the above equation we use the following identity
\begin{eqnarray}
&&\sum_{j(n)}(m_{j}\vec{v}_{jn}\vec{x}_{jn}).\nabla w(\vec{x}-\vec{x}_{n}) = \frac{d}{dt}\left[\mathbf{q}_{n}.\nabla w(\vec{x}-\vec{x}_{n})\right] + \nonumber
\\&&+\vec{v}_{n}\nabla.\nabla.[\mathbf{q}_{n}w(\vec{x}-\vec{x}_{n})]  +\nabla\times\nabla.\left[\mathbf{q}_{n}\times\vec{v}_{n}w(\vec{x}-\vec{x}_{n})\right] - \nonumber
\\ &&-\frac{1}{2}\nabla\times\sum_{j(n)}m_{j}(\vec{x}_{jn}\times\vec{v}_{jn}) w(\vec{x}-\vec{x}_{n}) ,\label{ident1}
\end{eqnarray}
so that we can write the average of current of matter of $n$-th local system (\ref{j0}) in the form
\begin{eqnarray}
<\vec{j}_{0}>_{n} &\approx&  \left<m_{n}\vec{v}_{n}\delta(\vec{x}-\vec{x}_{n})\right>  -\ederive{t}\nabla.\left<\mathbf{q}_{n}w(\vec{x}-\vec{x}_{n})\right> -\nonumber
\\&&-\nabla\times\nabla.\left<\mathbf{q}_{n}\times\vec{v}_{n}\delta(\vec{x}-\vec{x}_{n})\right>+ \nonumber
\\&&+\nabla\times\left<\vec{\mu}_{n}\delta(\vec{x}-\vec{x}_{n})\right> ,
\end{eqnarray}
where
\begin{equation}
 \vec{\mu}_{n} \equiv \frac{1}{2}\sum_{j(n)}m_{j}(\vec{x}_{jn}\times\vec{v}_{jn}),
\end{equation}
is the gravitomagnetic momentum of $n$-th local system.
Finally, according to (\ref{medj0} ), we can write
\begin{eqnarray}\label{j0final} 
\left<\vec{j}_{0}\right> &\approx& \vec{j}_{f} - \ederive{t}\nabla.\mathbf{Q} + \nonumber
\\&+&\nabla\times\left[\vec{M} -\nabla.\left<\sum_{n}\mathbf{q}_{n}\times\vec{v}_{n}\delta(\vec{x}-\vec{x}_{n})\right>\right], 
\end{eqnarray}
where we defined
\begin{equation}
 \vec{j}_{f} \equiv \left<\sum_{j(free)}m_{j}\vec{v}_{j}\delta(\vec{x} -\vec{x}_{j})\right> +\left<\sum_{n}m_{n}\vec{v}_{n}\delta(\vec{x}-\vec{x}_{n})\right> ,
\end{equation}
\begin{equation}
\vec{M} \equiv \left<\vec{\mu}_{n}\delta(\vec{x}-\vec{x}_{n})\right>,
\end{equation}
as the free current density and gravitomagnetization. The result is similar to that achieved in the electromagnetic case, with exception the polarization term, which vanish in the gravitational case.

\section*{Calculation of spatial average of $\rho_{k}$}
We will now calculate the spatial average of the term proportional to $ \rho_{k}$ which represents the contribution of kinetic energy. Similarly to what was done in the previous promediation we will write this term as
\begin{eqnarray}
\rho_{k} &=& \frac{1}{2}\sum_{j(free)}m_{j}v_{j}^{2}\delta(\vec{x}-\vec{x}_{j}) + \nonumber
\\&&+\frac{1}{2}\sum_{n}\sum_{k(n)}m_{k}\left(\vec{v}_{n}+\vec{v}_{kn}\right)^{2}\delta(\vec{x}-\vec{x}_{n}-\vec{x}_{kn}).\label{medrov2}
\end{eqnarray}
Again, the first term is the the kinetic energy of free bodies while the second is that of objects bounded in local systems. Let us look separately the contribution of the bounded term
\begin{eqnarray}
\left<\rho_{k}\right>_{n} &=& \frac{1}{2}\left<\sum_{k(n)}m_{k}\left(\vec{v}_{n}+\vec{v}_{kn}\right)^{2}\delta(\vec{x}-\vec{x}_{n}-\vec{x}_{kn})\right> \nonumber
\\ &\approx & \frac{1}{2}m_{n}v_{n}^{2}w(\vec{x} -\vec{x}_{n}) + \frac{1}{2}\nabla.\nabla.\mathbf{q}_{n}v_{n}^{2}w(\vec{x} -\vec{x}_{n}) -\nonumber
\\ &&- \vec{v}_{n}.\sum_{k(n)}[m_{k}\vec{v}_{kn}\vec{x}_{kn}].\nabla w(\vec{x} -\vec{x}_{n}) + \nonumber
\\&&+\frac{1}{2}\sum_{k(n)}m_{k}v_{kn}^{2}w(\vec{x} -\vec{x}_{n}).\label{rov2n}
\end{eqnarray}
To improve this expression we use the identity 
\begin{eqnarray}
&&\vec{v}_{n}.\sum_{k(n)}[m_{k}\vec{v}_{kn}\vec{x}_{kn}].\nabla w(\vec{x} -\vec{x}_{n}) = \nabla.\nabla.\left[\mathbf{q}_{n}v_{n}^{2} w(\vec{x} -\vec{x}_{n})\right] + \nonumber
\\&&\;\;+\frac{d}{dt}\nabla.\left[\mathbf{q}_{n}.\vec{v}_{n} w(\vec{x} -\vec{x}_{n})\right] +\nabla.\left[\vec{v}_{n}\times\vec{\mu}_{n}w(\vec{x} -\vec{x}_{n})\right]- \nonumber
\\&&\;\;- \nabla.\left[\mathbf{q}_{n}.\tot{\vec{v}_{n}}{t} w(\vec{x} -\vec{x}_{n})\right] +\nabla.\nabla.\left[\left(\mathbf{q}_{n}\times\vec{v}_{n}\right)\times\vec{v}_{n}w(\vec{x} -\vec{x}_{n})\right], 
\end{eqnarray}
to write the eq. (\ref{rov2n}) as
\begin{eqnarray}
\left<\rho_{k}\right>_{n} & \approx & \frac{1}{2}\left<m_{n}v_{n}^{2}\delta(\vec{x} -\vec{x}_{n})\right> -\frac{1}{2} \nabla.\nabla.\left<\mathbf{q}_{n}v_{n}^{2}\delta(\vec{x} -\vec{x}_{n})\right> +\nonumber
\\&&+\frac{1}{2}\left<\sum_{k(n)}m_{k}v_{kn}^{2}\delta(\vec{x} -\vec{x}_{n})\right> - \nonumber
\\ &&- \frac{\partial}{\partial t}\nabla.\left<\mathbf{q}_{n}.\vec{v}_{n} \delta(\vec{x}-\vec{x}_{n})\right> +\nabla.\left<\mathbf{q}_{n}.\tot{\vec{v}_{n}}{t}\delta(\vec{x}-\vec{x}_{n})\right> - \nonumber
\\ &&- \nabla.\nabla.\left<(\mathbf{q}_{n}\times\vec{v}_{n})\times\vec{v}_{n} \delta(\vec{x}-\vec{x}_{n})\right> -\nabla.\left<\vec{v}_{n}\times\vec{\mu}_{n}\delta(\vec{x}-\vec{x}_{n})\right>.
\end{eqnarray}
As each body follows the Newtonian equation of motion \cite{s.weinberg-gc}, we can write the following term as
\begin{eqnarray}
\left<\mathbf{q}_{n}.\tot{\vec{v}_{n}}{t} \delta(\vec{x}-\vec{x}_{n})\right> &\approx& \left<\mathbf{q}_{n}.\sum_{j(free)}Gm_{j}\frac{\vec{x}_{j}-\vec{x}_{n}}{\left| \vec{x}_{j}-\vec{x}_{n}\right|^{3}}\delta(\vec{x}-\vec{x}_{n})\right>+ \nonumber
\\ &&+ \left<\mathbf{q}_{n}.\sum_{m;m \neq n}Gm_{m}\frac{(\vec{x}_{m}-\vec{x}_{n})}{\left| \vec{x}_{m}-\vec{x}_{n}\right|^{3}}\delta(\vec{x}-\vec{x}_{n})\right>.
\end{eqnarray}
The first term represents the contribution of free bodies and can usually be ignored, but will be kept here for completeness, while the second term represents the contribution of local systems, which had been expanded in power series.
Finally we can return to equation (\ref{medrov2}) to obtain
\begin{equation}\label{rov2final} 
\left<\rho_{k}\right> \approx \left(\rho_{k}\right)_{f} -\nabla.\vec{P}_{k} -\nabla.\nabla.\mathbf{Q}_{k} , 
\end{equation}
where the free kinetic energy was defined as
\begin{eqnarray}
\left(\rho_{k}\right)_{f} &\equiv& \frac{1}{2}\left<\sum_{j(free)}m_{j}v_{j}^{2}\delta(\vec{x}-\vec{x}_{j})\right> +\frac{1}{2}\left<\sum_{n}m_{n}v_{n}^{2}\delta(\vec{x} -\vec{x}_{n})\right> +\nonumber
\\&&+\frac{1}{2}\left<\sum_{n}\sum_{k(n)}m_{k}v_{kn}^{2}\delta(\vec{x} -\vec{x}_{n})\right> ,
\end{eqnarray}
the kinetic polarization and the kinetic quadrupole momentum as
\begin{eqnarray} 
\vec{P}_{k} &=& \left<\sum_{n}\vec{v}_{n}\times\vec{\mu}_{n}\delta(\vec{x}-\vec{x}_{n})\right>
-\left<\smn\mathbf{q}_{n}.Gm_{m}\frac{(\vec{x}_{m}-\vec{x}_{n})}{\left|
\vec{x}_{m}-\vec{x}_{n}\right|^{3}}\delta(\vec{x}-\vec{x}_{n})\right> - \nonumber
\\ &&- \left<\sum_{n}\sum_{j(free)}\mathbf{q}_{n}.Gm_{j}\frac{\vec{x}_{j}-\vec{x}_{n}}{\left|
\vec{x}_{j}-\vec{x}_{n}\right|^{3}}\delta(\vec{x}-\vec{x}_{n})\right> + \nonumber
\\&&+\frac{\partial}{\partial t}\left<\sum_{n}\mathbf{q}_{n}.\vec{v}_{n}\delta(\vec{x}-\vec{x}_{n})\right>,
\\ \mathbf{Q}_{k} &=& \frac{1}{2}\left<\sum_{n}\mathbf{q}_{n}v_{n}^{2}\delta(\vec{x}
-\vec{x}_{n})\right>
+\left<\sum_{n}(\mathbf{q}_{n}\times\vec{v}_{n})\times\vec{v}_{n}
\delta(\vec{x}-\vec{x}_{n})\right>.
\end{eqnarray}
This term has no electromagnetic equivalent but behaves similarly to the average of density of matter. Unlike previous promediations that presents polarization terms.

\section*{Calculation of spatial average of $ \rho_{g}$}
Finally we calculate the average of term that represents the contribution of the Newtonian gravitational energy in source of gravitoelectric field. Using the coordinate system defined in the average of density of matter, FIG. \ref{fig: jjjj}, we can write it as
\begin{eqnarray}
\rho_{g} & = & \sum_{j(free)}m_{j}\delta(\vec{x} -\vec{x}_{j})\left[ \sum_{k(free), k\neq j}\frac{Gm_{k}}{\left|\vec{x} -\vec{x}_{k}\right|} +\sum_{n}\sum_{k(n)}\frac{Gm_{k}}{\left|\vec{x} -\vec{x}_{k}\right|}\right] + \nonumber
\\ &&+ \sum_{m}\sum_{j(m)}m_{j}\delta(\vec{x}- \vec{x}_{j})\left[\sum_{k(free)}\frac{Gm_{k}}{\left|\vec{x} -\vec{x}_{k}\right|} +\sum_{n}\!\!\!\!\!\sum_{\;\;\;\;\;\;\;\;k(n)_{_{_{\!\!\!\!\!\!\!\!\!\!\!\!\!\!\!\!\!\!\!\!k(n)\neq 
j(m)}}}}\frac{Gm_{k}}{\left|\vec{x} - \vec{x}_{k}\right|}\right] \nonumber
\\ &=& \sjkf\frac{Gm_{j}m_{k}}{\left|\vec{x}_{j} -\vec{x}_{k}\right|}\delta(\vec{x} - \vec{x}_{j}) +\nonumber
\\&&+\sum_{j(free)} \sum_{n}\sum_{k(n)}\frac{Gm_{k}m_{j}}{\left|\vec{x}_{j} -\vec{x}_{n}- \vec{x}_{kn}\right|}\delta(\vec{x} - \vec{x}_{j}) + \nonumber
\\ &&+ \sum_{k(free)}\sum_{m}\sum_{j(m)}\frac{Gm_{k}m_{j}}{\left|\vec{x}_{m} -\vec{x}_{k} +\vec{x}_{jm}\right|}\delta(\vec{x} - \vec{x}_{m} - \vec{x}_{jm}) + \nonumber
\\ &&+ \sum_{m,n}\sum_{\;\;j(m),k(n)_{_{_{\!\!\!\!\!\!\!\!\!\!\!\!\!\!\!\!\!\!\!\!\!\!\!\!\!\!\!k(n)\neq j(m)}}}}\frac{Gm_{j}m_{k}}{\left|\vec{x}_{m} - \vec{x}_{n} -(\vec{x}_{kn} -\vec{x}_{jm})\right|}\delta(\vec{x} - \vec{x}_{m} -\vec{x}_{jm}).
\end{eqnarray}
The first term above is the gravitational interaction of free bodies. The second and third is the interaction between free bodies and bounded bodies, while the last is the interaction between bounded bodies.
Again, the terms summed on the free bodies can be neglected, but we will keep these terms for completeness. Let us look separately the average of the last term
\begin{eqnarray}
<\rho_{g}>_{4} &=& \left<\sum_{m,n}\sum_{\;\;j(m),k(n)_{_{_{\!\!\!\!\!\!\!\!\!\!\!\!\!\!\!\!\!\!\!\!\!\!\!\!\!\!\!k(n)\neq j(m)}}}}\frac{Gm_{j}m_{k}}{\left|\vec{x}_{m} - \vec{x}_{n} -(\vec{x}_{kn} -\vec{x}_{jm})\right|}\delta(\vec{x} - \vec{x}_{m} -\vec{x}_{jm})\right> \nonumber
\\&\approx& \left<\smn\frac{Gm_{n}m_{m}}{\left|\vec{x}_{m} -\vec{x}_{n}\right|}\delta(\vec{x}-\vec{x}_{m})\right> +\nonumber
\\&&+\left<\sum_{n}\sjk\frac{Gm_{j}m_{k}}{\left|\vec{x}_{jn}-\vec{x}_{kn}\right|}\delta(\vec{x} -\vec{x}_{n})\right> + \nonumber
\\&& +\nabla.\left< 2\smn\mathbf{q}_{m}.\frac{Gm_{n}(\vec{x}_{m}-\vec{x}_{n})}{\left|\vec{x}_{m} - \vec{x}_{n}\right|^{3}}\delta(\vec{x}-\vec{x}_{m})\right> -\nonumber
\\&&-\nabla.\left<\sum_{n}\sjk\frac{Gm_{k}m_{j}\vec{x}_{jn}}{\left|\vec{x}_{jn}-\vec{x}_{kn}\right|}\delta(\vec{x} -\vec{x}_{n})\right> +\nonumber
\\&&+\nabla.\nabla.\left<\frac{1}{2}\sum_{n}\sjk\frac{Gm_{k}m_{j}\vec{x}_{jn}\vec{x}_{jn}}{\left|\vec{x}_{jn}-\vec{x}_{kn}\right|}\delta(\vec{x} -\vec{x}_{n})\right> +\nonumber
\\&&+\nabla.\nabla.\left<\smn\frac{Gm_{n}\mathbf{q}_{m}}{\left|\vec{x}_{m} -\vec{x}_{n}\right|}\delta(\vec{x}-\vec{x}_{m})\right> .
\end{eqnarray}
where in the last passage we expand $ \left| \vec{x}_{m} - \vec{x}_{n} - (\vec{x}_{kn} - \vec{x}_{jm})\right|^{-1}$ around $ \vec{x}_{kn} - \vec{x}_{jm} = \vec{0} $, for $ m \neq n $, $ w (\vec{x} - \vec{x}_{m} - \vec{x}_{jm}) $ around $ \vec{x}_{jm} = \vec{0} $ and we eliminated higher order terms.
Now we focus on the average of the third term of $ \rho_{g} $, that is
\begin{eqnarray}
<\rho_{g}>_{3} &=&\left<\sum_{k(free)}\sum_{m}\sum_{j(m)}\frac{Gm_{k}m_{j}}{\left|\vec{x}_{m} -\vec{x}_{k} +\vec{x}_{jm}\right|}\delta(\vec{x} - \vec{x}_{m} - \vec{x}_{jm})\right> \nonumber
\\ &\approx & \left<\sum_{k(free)}\sum_{m}\sum_{j(m)}\frac{Gm_{k}m_{j}}{\left|
\vec{x}_{j}-\vec{x}_{k}\right|}\delta(\vec{x}-\vec{x}_{m})\right> -\nonumber
\\&&-\nabla.\left<\sum_{k(free)}\sum_{m}\sum_{j(m)}\frac{Gm_{k}m_{j}\vec{x}_{jm}}{\left|
\vec{x}_{j}-\vec{x}_{k}\right|^{2}}\delta(\vec{x}-\vec{x}_{m})\right>+ \nonumber
\\&&+ \nabla.\nabla.\left<\frac{1}{2}\sum_{k(free)}\sum_{m}\sum_{j(m)}\frac{Gm_{k}m_{j}\vec{x}_{jm}\vec{x}_{jm}}{\left|\vec{x}_{j}-\vec{x}_{k}\right|}\delta(\vec{x}-\vec{x}_{m})\right>.
\end{eqnarray}
Finally, adding the contribution of all terms we can write the average of $ \rho_{g} $ as
\begin{equation}
 <\rho_{g}> \approx (\rho_{g})_{f} +\nabla.\vec{P}_{g} +\nabla.\nabla.\mathbf{Q}_{g} ,
\end{equation}
where we define the free gravitational energy
\begin{eqnarray}
(\rho_{g})_{f} &=& \left<\sjkf\frac{Gm_{j}m_{k}}{\left|\vec{x}_{j} -\vec{x}_{k}\right|}\delta(\vec{x} - \vec{x}_{j})\right> +\left<\smn\frac{Gm_{n}m_{m}}{\left|\vec{x}_{m} -
\vec{x}_{n}\right|}\delta(\vec{x}-\vec{x}_{m})\right>+\nonumber
\\&&+\left<\sum_{j(free)}\sum_{n}\sum_{k(n)}\frac{Gm_{k}m_{j}}{\left|\vec{x}_{j} -
\vec{x}_{k}\right|}\delta(\vec{x} - \vec{x}_{j})\right> + \nonumber
\\ &&+ \left<\sum_{k(free)}\sum_{m}\sum_{j(m)}\frac{Gm_{k}m_{j}}{\left|
\vec{x}_{j}-\vec{x}_{k}\right|}\delta(\vec{x}-\vec{x}_{m})\right> +\nonumber
\\&& +\left<\sum_{n}\sjk\frac{Gm_{j}m_{k}}{\left|\vec{x}_{jn}-\vec{x}_{kn}\right|}\delta(\vec{x}
-\vec{x}_{n})\right>,
\end{eqnarray}
and the quantity
\begin{eqnarray}
\vec{P}_{g} &=& 2\left<\smn\mathbf{q}_{m}.\frac{Gm_{n}(\vec{x}_{m}-\vec{x}_{n})}{\left|\vec{x}_{m} -\vec{x}_{n}\right|^{3}}\delta(\vec{x}-\vec{x}_{m})\right>-\nonumber
\\&&-\left<\sum_{n}\sjk\frac{Gm_{k}m_{j}\vec{x}_{jn}}{\left|\vec{x}_{jn}-\vec{x}_{kn}\right|}\delta(\vec{x}-\vec{x}_{n})\right> - \nonumber
\\&&-\left<\sum_{k(free)}\sum_{m}\sum_{j(m)}\frac{Gm_{k}m_{j}\vec{x}_{jm}}{\left|
\vec{x}_{j}-\vec{x}_{k}\right|^{2}}\delta(\vec{x}-\vec{x}_{m})\right>,
\\ \mathbf{Q}_{g} &=& \left<\frac{1}{2}\sum_{k(free)}\sum_{m}\sum_{j(m)}\frac{Gm_{k}m_{j}\vec{x}_{jm}\vec{x}_{jm}}{\left|\vec{x}_{j}-\vec{x}_{k}\right|}\delta(\vec{x}-\vec{x}_{m})\right> +\nonumber
\\&&+ \left<\smn\frac{Gm_{n}\mathbf{q}_{m}}{\left|\vec{x}_{m} -\vec{x}_{n}\right|}\delta(\vec{x}-\vec{x}_{m})\right> + \nonumber
\\ &&+ \left<\frac{1}{2}\sum_{n}\sjk\frac{Gm_{k}m_{j}\vec{x}_{jn}\vec{x}_{jn}}{\left|\vec{x}_{jn}-\vec{x}_{kn}\right|}\delta(\vec{x} -\vec{x}_{n})\right>,
\end{eqnarray}
are the gravitational polarization and the quadrupole momentum of gravity. Again, despite having no electromagnetic corresponding, the average this term exhibits behaviour similar to other sources of gravitoelectric field, including polarization terms.
\section{Definition of Large Scale Fields}
In the previous sections we calculated the averages needed to write the promediated Maxwell-Einstein's equation. Finally we find the result
\begin{eqnarray}
&&\nabla.\vec{E}_{g} \approx -\frac{1}{\varepsilon_{g}}\left[\rho_{f} +3\left(\rho_{k}\right)_{f} +(\rho_{g})_{f} +\right.\nonumber
\\&&\;\;\;\;\;\;\;\;\;\;\;\;\;\;+\left.\nabla.\left[\vec{P}_{g} -3\vec{P}_{k} +\nabla.\left(\textbf{Q} -3\mathbf{Q}_{k} +\mathbf{Q}_{g}\right)\right]\right] ,\label{promed1}
\\&& \nabla.\vec{B}_{g} = 0 ,
\\&& \nabla\times\vec{E}_{g} = -\derive{\vec{B}_{g}}{t} ,
\\&& \nabla\times\vec{B}_{g} \approx -\mu_{g}\vec{j}_{f} +\mu_{g}\ederive{t}\nabla.\mathbf{Q} -\nonumber
\\&&\;\;\;\;\;\;\;\;\;\;\;\;\;\;\;\;-\mu_{g}\nabla\times\left[\vec{M} -\nabla.\left<\sum_{n}\mathbf{q}_{n}\times\vec{v}_{n}\delta(\vec{x}-\vec{x}_{n})\right>\right] +\derive{\vec{E}_{g}}{t} .\label{promed2}
\end{eqnarray} 
The equation (\ref{promed1}) leads us to define the field and the effective density
\begin{eqnarray}
\vec{D}_{g} &\approx&  \vec{E}_{g} +\frac{1}{\varepsilon_{g}}\left[\vec{P}_{g} -3\vec{P}_{k} +\nabla.\left(\textbf{Q} -3\mathbf{Q}_{k} +\mathbf{Q}_{g}\right)\right], \label{defD}
\\ \rho_{e} &=& \rho_{f} +3\left(\rho_{k}\right)_{f} +(\rho_{g})_{f},
\end{eqnarray}
in order to validate the equation
\begin{equation}
 \nabla.\vec{D}_{g} = -\frac{1}{\varepsilon_{g}}\rho_{e} .
\end{equation}
The definition (\ref{defD}) allows us to write eq. (\ref{promed2} ) in the form
\begin{equation}
\nabla\times\vec{H}_{g} = -\mu_{g}\vec{j}_{e} +\derive{\vec{D}_{g}}{t},
\end{equation}
where we define the effective current and the large scale field
\begin{eqnarray}
&&\vec{j}_{e} \approx \vec{j}_{f} +\ederive{t}\left[ \vec{P}_{g} -3\vec{P}_{k}  +\nabla.\left(\mathbf{Q}_{g} -3\mathbf{Q}_{k}\right)\right].
\\&&\vec{H}_{g} \approx \vec{B}_{g} +\mu_{g}\nabla\times\left[\vec{M} -\nabla.\left<\sum_{n}\mathbf{q}_{n}\times\vec{v}_{n}\delta(\vec{x}-\vec{x}_{n})\right>\right].
\end{eqnarray}
According to the above definitions, we can write the large scale Maxwell-Einstein's equation in the form
\begin{eqnarray}
\nabla.\vec{D}_{g} &=& -\frac{1}{\varepsilon_{g}}\rho_{e},
\\ \nabla.\vec{B}_{g} &=& 0 ,
\\ \nabla\times\vec{E}_{g} &=& -\derive{\vec{B}_{g}}{t} ,
\\ \nabla\times\vec{H}_{g} &=& -\mu_{g}\vec{j}_{e} +\derive{\vec{D}_{g}}{t} .
\end{eqnarray}
Finally, we can note that both the free and the effective mass are conserved, i.e. 
\begin{eqnarray}
0 &=& \nabla.\vec{j}_{f} +\derive{\rho_{f}}{t},
\\0 &=&  \nabla.\vec{j}_{e} +\derive{\rho_{e}}{t}.\label{cmef}
\end{eqnarray}
The conservation law (\ref{cmef}) indicates that the effective current is a current associated with source of gravitoelectric field. This indicates a
symmetry in large scale equations much greater than in fundamental equations (\ref{m-e1}) - (\ref{m-e4}). In latter the source of 
gravitomagnetic field was only the flow of matter, without regard the flow of other forms of energy, whereas the source of the
gravitoelectric field take them into account. In large scale equations that asymmetry was broken, because the source of the field $ \vec{H}_{g} $ is the same order flow of the source of field $\vec{D}_{g}$.

\section{Conclusions and Perspectives}
We show in this paper that it is possible to construct fields with independent elements of the metric in the fourth order post-Newtonian
approximation, which obey equations similar to Maxwell's electromagnetism. Through these equations becomes explicit that not only the matter interacts gravitationally also more energy - in this case the gravitational
potential and kinetic energy. Due to gauge obeyed by these fields, from fixing coordinates, the propagation of these fields occurs at the speed
of light, which agrees with the spirit of special relativity. Despite this progress in building an analogy between both theories, it is unable to write the equation of motion in the form of the Lorentz equation.
With gravitoelectromagnetic fields we found the effective fields for a distribution of matter organized in local systems, as occurs in the universe where matter is organized into bounded structures as solar systems, on a scale, galaxies,
on a scale greater. Only after this last level is that we can use a cosmological model, as Friedmann, which presupposes matter homogeneously distributed.
The equations for these effective fields, $D_{g} $ and $H_{g}$, exhibited a greater symmetry than the equations of fundamental fields, since their sources satisfy an equation of continuity in the same order, i.e., the source of field $ H_{g} $ is the flow associated with the source of field $ D_{g}$.
We show that the definition of these fields are greatly simplified by choosing the coordinate system fixed at the center of mass of local systems, where the terms of dipole vanish.
As perspectives to further studies to be made based on the results obtained in this work we can mention the inclusion of the cosmological constant, as well as calculate the corrections of order higher than those considered in this work.

\section*{Acknowledgments}

We acknowledge the financial support provided by Funda\c c\~ao Cearense de Apoio ao Desenvolvimento Cient\'\i fico e Tecnol\'ogico (FUNCAP), the Conselho Nacional de 
Desenvolvimento Cient\'\i fico e Tecnol\'ogico (CNPq) and FUNCAP/CNPq/PRONEX.
%\bibliographystyle{unsrt}
%\bibliography{artigos,book}

\end{document}